\documentclass[aip, pop, amsfonts, amssymb, amsmath ,reprint,subscriptadress]{revtex4-1}
\usepackage{times}
\usepackage{bm}
\usepackage{hyperref}
\usepackage{graphicx}
\usepackage{epsfig} 
\usepackage{epstopdf}
\usepackage{color}

\begin{document}

\title{Applications of Large Eddy Simulation methods to gyrokinetic turbulence}
%,
%
\author{A.~Ba\~n\'on Navarro}
\email{abanonna@ipp.mpg.de}
\affiliation{Max-Planck-Institut f\"ur Plasmaphysik, EURATOM Association, D-85748 Garching, Germany}

\author{B.~Teaca}
\affiliation{Applied Mathematics Research Centre, Coventry University, Coventry CV1 5FB, United Kingdom}
\affiliation{Max-Planck f\"ur Sonnensystemforschung, Max-Planck-Str.~2, D-37191 Katlenburg-Lindau, Germany}
\affiliation{Max-Planck/Princeton Center for Plasma Physics}

\author{F.~Jenko}
\affiliation{Max-Planck-Institut f\"ur Plasmaphysik, EURATOM Association, D-85748 Garching, Germany}
\affiliation{Max-Planck/Princeton Center for Plasma Physics}

\author{G.~W.~Hammett}
\affiliation{Max-Planck/Princeton Center for Plasma Physics}
\affiliation{Princeton Plasma Physics Laboratory, Princeton University, P.O. Box 451, Princeton, New Jersey 08543, USA}

\author{T.~Happel}
\affiliation{Max-Planck-Institut f\"ur Plasmaphysik, EURATOM Association, D-85748 Garching, Germany}

\author{the ASDEX Upgrade Team}

\date{\today}

\begin{abstract}
The Large Eddy Simulation (LES) approach -  solving  numerically  the large scales of a turbulent system and accounting for the small-scale influence through a model -   is applied  to nonlinear gyrokinetic systems that are driven by a number of different microinstabilities.  Comparisons between modeled, lower resolution, and higher resolution  simulations are  performed for  an experimental measurable quantity,   the electron density fluctuation spectrum.   Moreover, the validation and applicability of LES is demonstrated through a series of diagnostics based on the free energetics of the system.
\end{abstract}
% Gyrokinetics in plasmas; Plasma turbulence; Gyrofluid and gyrokinetic plasma simulations,
%
%
\maketitle

%%%%%%%%%%%%%%%%%%%%%%%%%%%%%%%%%%%%%%%%%
%%%%%%%%%%%%%%%%%%  Introduction %%%%%%%%%%%%%%%%
%%%%%%%%%%%%%%%%%%%%%%%%%%%%%%%%%%%%%%%%%

%%%%%%%%%%%%%
\section{Introduction}

Large Eddy Simulation (LES)  methods were first introduced within the computational  fluid dynamics  community\cite{smagorinsky-MWR-1963} in an attempt to focus on the large scales of a turbulent flow, which often contain the information of interest for practical applications, using the least possible amount of computational resources. Simpler versions of  LES methods are based on a phenomenological approach to turbulence, which, for a fluid described by the Navier-Stokes equations, can be understood in terms of two concepts, scale separation and redistribution of energy between  scales. Indeed, the Reynolds number, which is used to characterize different flow regimes of a fluid, measures the ratio between the forcing scale and the dissipation scale in the system. For weakly turbulent flows (or equivalently for low Reynolds numbers), the small separation of scales implies the excitation of only a few degrees of freedom. As one approaches a fully developed turbulent state, the scale separation increases, and more degrees of freedom  become excited. As the forcing and dissipation start to act primarily at completely different scales, an inertial range develops to bridge the two effects. The inertial range, dynamically dominated by the nonlinear  couplings, serves to redistribute the energy from the large forcing scale to the small dissipation scale, in a process known as a cascade~\cite{k41}.  This redistribution of energy is expected to have a universal character and leads to the development of  power laws for certain spectral quantities. As such, accurately recovering the correct power law exponents is a sign of an adequately resolved simulation. 

Turbulence in magnetized plasmas is more complex than fluid turbulence since it involves multi-field dynamics, important kinetic effects, and the possibility to dissipate energy at different (phase space) scales. Moreover, plasma turbulence can be driven by a large variety of different microinstabilities - including ion temperature gradient (ITG) modes, trapped electron modes (TEMs), and electron temperature gradient (ETG) modes, which may differ significantly in their characteristic spatio-temporal scales as well as in their fluctuation power law spectra. In particular,  in the context of plasma turbulence described by the gyrokinetic (GK) model~\cite{brizard-RMP-2007, sugama00, abel13}, several theories try to explain the  power laws found in experiments or in direct numerical simulations  by means of concepts like nonlinear phase mixing~\cite{scheko08,plunk10}, critical balance~\cite{barnes11}, or damped eigenmodes~\cite{hatch11}. For this reason, a correct identification of the power law exponents is important for the understanding of the underlying physics and useful for providing constraints  for simple physical models. From an experimental point of view, the knowledge of characteristic scales and wavenumber spectra is important for the clear identification of the different turbulence regimes, in which various microinstabilities can affect the confinement of particles and heat in different ways~\cite{conway08}. With the recent improvements in fluctuation diagnostics, such as the new Doppler reflectometer in the ASDEX Upgrade tokamak~\cite{tim13_a}, it is now possible to measure turbulence characteristics with higher precision, allowing for better direct comparisons between the experimental data and the results of nonlinear gyrokinetic simulations.

Unfortunately, within the context of nonlinear gyrokinetics, ensuring that all of the relevant phase-space dissipation mechanisms are adequately resolved (so that the fluctuation statistics are adequately described) can be very expensive from a computational point of view~\cite{goerler08_2}. Hence, the LES technique used for simulation fluid turbulence has been applied to plasma turbulence, first using simpler shearing-rate based sub-grid models in gyrofluid\cite{Smith97} and gyrokinetic\cite{Belli2008} simulations, and recently using more advanced dynamic sub-grid models\cite{morel11,morel13}. The same ideas, to resolve the largest scales in the system and model the influence of small ones, are applied to the gyrokinetic equations and give rise to the Gyrokinetic Large Eddy Simulation (GyroLES) approach. Previous efforts in this new field were focused on saving the computational time as much as possible, while having the most accurate possible results in terms of global transport quantities, such as cross-field heat and particle fluxes.  This requires  retaining only relatively few scales of motion, and simulations speedups by factors of $20-30$ have been achieved. The present paper is not aimed at calculating only global transport quantities at minimal computational cost, but to demonstrate that the GyroLES approach, using a similar resolution that is used in present simulations, yields more accurate power law exponents  for different quantities  and  for a wide range of parameters and  instabilities, at a lower computational cost.   In particular,  we will show that to have at least the same accuracy in the resulting power laws  as in a GyroLES simulation, one will need to perform a simulation with at  least two times more resolution in both perpendicular spatial directions.

The remainder of this paper is organized as follows. The gyrokinetic model is briefly introduced in Sec.~\ref{sec:GK}, and the GyroLES approach is summarized in Sec.~\ref{sec:gyroLES}. Numerical results are then presented in Sec.~\ref{sec:numerics}. Here, a description of the different cases and instabilities is provided, followed  by an analysis of the performance of the GyroLES methods. The latter will be focused on  an experimentally accessible quantity,  namely  the electron  density fluctuation  spectrum.  Moreover, in order to better understand the range of applicability of the GyroLES approach for the different cases,  several  diagnostics based on the free energy of the system  will  be  introduced and analyzed in  detail in the last part of this section.   Finally, conclusions and  discussions of the main results will be  given.
  
%%%%%%%%%%%%%%%%%%%%%%%%%%%%%%%%%%%%%%%%%%%%%%%%%%%%%%%%%%%%
%%%%%%%%%%%%%%%%%%%%%%%%%%%%%%%%%%%%%%%%%%%%%%%%%%%%%%%%%%%%
%  The gyrokinetic equations
%%%%%%%%%%%%%%%%%%%%%%%%%%%%%%%%%%%%%%%%%%%%%%%%%%%%%%%%%%%%
\section{Gyrokinetic model} \label{sec:GK}

The simulations presented below are performed with  the gyrokinetic code {\sc Gene}~\cite{gene}. It integrates in time ($t$) the nonlinear gyrokinetic equations on a fixed grid that discretizes the five-dimensional phase space. {\sc Gene} uses a field aligned coordinate system that exploits the scale separation between the perpendicular and parallel directions. 
The real space non-orthogonal coordinates are represented by ${\{x,y,z\}}$,  where $z$ is the coordinate along the magnetic field line, while  the radial coordinate $x$ and the binormal coordinate $y$ are orthogonal to the magnetic field. The velocity space coordinates ${ \{v_{\parallel}, \mu \} }$ are, respectively, the velocity parallel to the magnetic field and the magnetic moment.  For simplicity, we restrict ourselves here to the local approximation,  although {\sc Gene} can also be used as a global code~\cite{goerler11}. In this case, the coordinates  perpendicular to the magnetic field are Fourier transformed $(x,y) \rightarrow (k_x,k_y)$. Symbolically, the evolution equation for the distribution function $g_{j} = g_{j}(k_x, k_y, z, v_\parallel, \mu, t)$ can be expressed as
\begin{align}
\frac{ \partial  g_{j}} {\partial t} =  L [g_{j}] + D[g_{j}] + N[g_{j}]  \;.
\label{eq:GK}
\end{align}
 Typically, the index $j$ takes  two values, $j = i$ for the ions and $j = e$ for the electrons. 
  
The first term in Eq.~\eqref{eq:GK} is a linear term which  can be split into three contributions, $L[g_{j}] = G[g_{j}] + L_C[g_{j}]+ L_\parallel[g_{j}]$. Here,  $G$  represents the influence of the density and temperature gradients, $L_C$ describes effects due to magnetic curvature, and $L_\parallel$ contain the parallel dynamics involving magnetic trapping as well as  linear Landau damping. The next term in  Eq.~\eqref{eq:GK} is the dissipation term, $D[g_{j}]$, which is represented by a Landau-Boltzmann collision operator or by fourth-order hyper diffusion operators in the collisionless case. Finally, $N[g_{j}]$ is the nonlinear term, 
\begin{align}
N[g_{j}]= \frac{\partial \chi}{\partial y}\frac{\partial h_j}{\partial x} - \frac{\partial \chi}{\partial x}\frac{\partial h_j}{\partial y}\;,
\label{nonlin}
\end{align}
where $h_j$ is the non-adiabatic part of the perturbed distribution functions and $\chi$ are the electrodynamic field contributions, obtained self-consistently from the Poisson-Amp\`ere laws for gyrokinetics. The nonlinear term has the fundamental role of coupling different
scales in phase space and leads to an effective coupling of perpendicular $k_x$ and $k_y$ modes. For the explicit form of the linear terms see Ref.~[\onlinecite{banon11}], although the knowledge of their explicit form in not necessary for the understanding of the current paper. 

%%%%%%%%%%%%%%%%%%%%%%%%%%%%%%%%%%%%%%%%%%%%%%%%%%%%%%%%%%%%
%%%%%%%%%%%%%%%%%%%%%%%%%%%%%%%%%%%%%%%%%%%%%%%%%%%%%%%%%%%%
%  The Filtered gyrokinetic equation
%%%%%%%%%%%%%%%%%%%%%%%%%%%%%%%%%%%%%%%%%%%%%%%%%%%%%%%%%%%%
\section{The Filtered gyrokinetic equation} \label{sec:gyroLES}

Large Eddy Simulations for gyrokinetics require a separation between the large (resolved) and the small scales in the system. As we are only interested in a separation of perpendicular spatial scales, characterized by modes in $k$ space, we introduce a cutoff wavenumber $k_c$ that separates the two.
Omitting the functional dependences of the terms and the distribution function's species label, the evolution
equation for the large scales ($|k_x| \le k_c$ and $|k_y| \le k_c$) can be written as
\begin{align}
\frac{  \partial} {\partial t} g^{<}_{k | k_c}  =  L^{<}_{k | k_c}  + D^{<}_{k | k_c} + N^{<}_{k | k_c} + N^{SGS}_{k|k_c},
\label{eq:LES}
\end{align}
where the subscript notation ${k | k_c}$ indicates that the $k$ dependent terms have been parametrized with respect to the cutoff wavenumber $k_c$.  In addition, the $<$ superscript notation indicates that in computing the large scale terms, only modes satisfying the inequality $k \le k_c$ are retained. This is always true for the linear terms. However, as the nonlinear term $N_k$ for the large scales ($k \le k_c$) mixes the large and the small scales, we split its contribution into two parts. One part, $N^{<}_{k | k_c}$, that contains interactions occurring only between large scale modes  and another part that takes into account the interactions with the small sub-grid scale (SGS) modes, for which $k > k_c$. The sub-grid term $N^{SGS}_{k|k_c}$ is the only term that cannot be expressed as a function of solely the resolved scales $k \le k_c$. Taking into account that Eq.~\eqref{eq:LES} is just the GK equation rewritten for modes $k < k_c$,   the sub-grid term is simply
\begin{align}
N^{SGS}_{k|k_c} = N_k -  N^{<}_{k | k_c}\;. 
\end{align}
The GyroLES approach consists in replacing this SGS term by a good model, which only depends on the resolved quantities $g^{<}_{k | k_c}$ and a set of free parameters $a = \{a_1, a_2, ...\}$,
\begin{align}
\label{eq:subgridterm} 
N^{SGS}_{k|k_c}   \approx  M^{<}_{k | k_c}[a].
\end{align} 
The free parameters must then be calibrated appropriately.

Through a process known as the {\it dynamic procedure}, it is possible to calibrate automatically all free parameters in the model. In a first step, the procedure requires the introduction of an additional cutoff scale $k_t$, with $k_t < k_c$, known as a test-scale. The resulting test-filtered gyrokinetic equation,
\begin{align}
\frac{  \partial} {\partial t}   g^{<}_{k | k_t} \! \!= \! L^{<}_{k | k_t}  \!+\! D^{<}_{k | k_t} \!+\! N^{<}_{k | k_t} \!+\! N^{STS}_{k | k_t,k_c}\!+\!  M^{<}_{k | k_c}[a]\,,
\label{eq:TEST_1}
\end{align}
contain the sub-test-scales (STS) term, parametrized in respect to $k_t$ and $k_c$. Since $N^{STS}_{k|k_t,k_{c}}=N^{<}_{k | k_c}-N^{<}_{k | k_t}$, it can be computed explicitly as resolved scales up to $k_c$ are known. In a second step, a cutoff wavenumber $k_t$ is introduced directly into Eq.~\eqref{eq:GK}. This yields (for scales $k < k_t$)
\begin{eqnarray}
\frac{  \partial} {\partial t}   g^{<}_{k | k_t}  &=&  L^{<}_{k | k_t} + D^{<}_{k | k_t}  + N^{<}_{k | k_t} +  N^{SGS}_{k|k_t} \nonumber \\
&=& L^{<}_{k | k_t} + D^{<}_{k | k_t} + N^{<}_{k | k_t} +  M^{<}_{k | k_t}[a]\;,
\label{eq:TEST_2}
\end{eqnarray}
where in the last equation, the sub-grid term has been replaced by the same model as in Eq.~\eqref{eq:TEST_1}. Although the same free parameters $\{a_1, a_2, ...\}$ are used, this models acts now in a more limited simulation ($k_t < k_c$), and therefore, its amplitudes will be adjusted accordingly. Equating Eqs.~\eqref{eq:TEST_1} and~\eqref{eq:TEST_2}, up to test scales $k \le k_t$ yields  an identity for the sub-test-scale term and the model, known as the Germano identity~\cite{germano91, germano92},
\begin{align}
N^{STS}_{k | k_t, k_c} + M^{<}_{k | k_c}[a] -  M^{<}_{k | k_t}[a] = 0 \quad \mbox{for} \quad k \le k_t\;.
\end{align}
The unknowns of the Germano identity, i.e., the free parameters of the model $\{a_1, a_2, ...\}$, can then be calculated by an optimization of this difference with respect to the unknowns (least squares method),
\begin{align}
&\frac{\partial}{\partial \{{a_1, a_2, ...\}}}\!  \left<\!\! \left ( N^{STS}_{k | k_t,k_c} \!+\! M^{<}_{k | k_c}[a] \!-\!  M^{<}_{k | k_t}[a] \!  \right )^2 \right >_{ {\Lambda}}\!\! \! \! \! \!  = 0 \,,
\label{eq:par}
\end{align}
where $\left< \dots \right>_{\Lambda}$ represents phase space ($\Lambda$) integration.
Note that,  if  more  than  one kinetic species are  being solved, the resulting parameters of the model are  species dependent. This allows one  to separately  model  the different  species in the system. 

The numerical resolution used in a code is indicated by introducing a cutoff filter denoted by
a cutoff filter denoted by $\overline{\cdots}$, with a characteristic length $\overline{\ell}  = 1 /k_c$. This filter  sets to zero the smallest scales in the distribution function $g_j$, characterized by all modes larger than   $ k > k_c$.  In particular, for the GK equation solved by the {\sc Gene} code, the cutoff $k_c$ is performed in the perpendicular plane and the filter is implemented  numerically by reducing the number of grid points in ($k_x, k_y$) space. 

In previous works\cite{morel11,morel13}, a hyper-diffusion model for the sub-grid term was proposed,
\begin{align}
M[\bar{g}, a_x, a_y] = -(a_x \bar{\ell}_x^\alpha (k_x \bar{\ell}_x)^n 
   + a_y \bar{\ell}_y^\alpha (k_y \bar{\ell}_y)^n ) \bar{h}
\label{eq-sgs-model}
\end{align}
with $n = 4$. Here, ${\bar{\ell}_x, \bar{\ell}_y}$ represents the characteristic filter scale in the perpendicular directions, 
${a_x, a_y}$ are the free
parameters, and $h$ is the non-adiabatic part of the perturbed
distribution function.  (As pointed out in Ref.~[\onlinecite{Catto1978}], renormalized damping models should only damp the non-adiabatic part of the distribution function.  The adiabatic/Boltzmann part is already in a state of maximum entropy, so reducing it would reduce the entropy.  Furthermore, the nonlinearity vanishes on the adiabatic part of the distribution function.)  Note that the damping rate in each direction in the sub-grid term, $a_i \bar{\ell}_i^\alpha$, has units of $1/t$.  In those previous works, we used a dimensional analysis based on the ``free energy flux density'' to fix the cutoff scale
exponent, which gives $\alpha = 1/3$. However, here we will
use a somewhat more conventional estimate based on the free energy flux and analogies to standard fluid turbulence, which leads to $\alpha = -2/3$.  In the Kolmogorov picture of fluid turbulence, quantities at scale $\ell$ in the inertial range can depend only on the scale $\ell$ and on the energy flux $\epsilon$ (in a plasma, the related quantity is the free energy flux), which has units of $\epsilon \sim v_\ell^3 / \ell$, where $v^2_\ell$ is the energy per unit mass in eddies of scale $\ell$.  (See for example Sec. 7.2 of Ref.~[\onlinecite{Frisch1995}].)  Dimensional analysis shows that if the damping rate $a_i \bar{\ell}_i^\alpha$ can depend only on these two parameters, $a_i \bar{\ell}_i^\alpha \sim \epsilon^\beta \bar{\ell}_i^\gamma$, then that means that $\alpha = \gamma = -{2/3}$, and $\beta = 1/3$, so that $a_i \sim \epsilon^{1/3}$.  Physically, this scaling of the sub-grid model means that the damping rate scales with the eddy turnover rate, which increases at smaller scales like $\ell^{-2/3}$ in the inertial range.  

This $\ell^{-2/3}$ scaling is tied to the inertial range energy spectrum of fluid turbulence of $E_k \sim k^{-5/3}$.  However, there are additional parameters that may affect plasma turbulence so that different spectral slopes may be seen in different types of plasma turbulence (as we will see in this paper), and thus the optimal value of the scaling with $\ell$ might change.  This is because of several factors, including the anisotropy and additional modes in plasmas.  I.e., the energy cascade rate in the perpendicular directions can be affected by energy cascades to finer scales in the parallel direction, to finer scales in velocity space, and by coupling to modes at the same spatial scale that are Landau damped\cite{hatch11}.  Also, the relation between the (free) energy flux and the eddy velocity spectrum in plasmas is not as straightforward as it is neutral fluids because of finite-Larmor-radius and other effects.  

For cases where the coefficients $a_i$ are fit (using the procedure defined below) with a test filter width that is a factor of 2 larger than the resolved scale, the new scaling of $M \sim a_i \bar{\ell_i}^{-2/3}$ would make $M$ a factor of 2 larger, if the coefficient $a_i$ was the same.  In fact, because of anisotropies and nonlinearities in plasma dynamics, we have found that this change in the exponent of $\bar{\ell}$ causes $a_i$ to also increase, so that the overall increase in the magnitude of sub-grid model damping rate can be a factor of $\sim 5$ larger in some cases.  Our general experience is that this stronger value of sub-grid damping rate has made it more robust and effective.  In general, it seems better if the coefficient of the sub-grid term is somewhat larger than optimal instead of too small.  Because of the hyperdiffusion form $\sim k^4$ of the sub-grid term, if the damping rate is too strong at the grid scale $k \sim 1/\bar{\ell}$, it will be about right at a somewhat smaller value of $k$.  But if the damping rate is too weak at the perpendicular grid scale, then there will be a bottleneck for energy cascade in the perpendicular direction, and energy transfers will instead be forced in the parallel direction or to other modes with stronger Landau damping, so that the spectra are more strongly distorted.

The resulting filtered gyrokinetic equation solved in {\sc Gene} then reads
\begin{align}
\frac{ \partial \overline{g} } {\partial t} &= L [\overline{g}] + D [\overline{g}]  + N [\overline{g}]   + N_{\overline{\ell}, {\ell}} \nonumber \\
&= L [\overline{g}]  + D [\overline{g}]+ N [\overline{g}]   + M[\overline{g}, a_x, a_y].
\end{align}
Here, $N_{\overline{\ell}, {\ell}}$ is the sub-grid term, which for clarity of the presentation, is represented by a notation that indicates  the fact that this term contains the influence of both the resolved scales $\overline{\ell}$ and the sub-grid scales  ${\ell}$.

The parameters  $\{ a_x , a_y \}$  can now be calculated with the application of the dynamic procedure by introducing the additional test-filter denoted in the following by $\widehat{\cdots}$, with a characteristic  length  taken simply as $\widehat{\ell}=2/k_c = 2 \, \overline{\ell}$. The resulting optimization of the system of  equations given by Eq.~\eqref{eq:par} yields
\begin{align}
a_x = & \frac{1}{\gamma_x}  \frac{\left < m_x N_{\widehat{\ell}, \overline{\ell}} \right >_{\Lambda} \left < m_y^2 \right >_{\Lambda} - \left < m_y N_{\widehat{\ell}, \overline{\ell}} \right >_{\Lambda} \left < m_y m_x \right >_{\Lambda}}{\left < m_x m_y \right >_{\Lambda}^2 - \left < m_x^2 \right >_{\Lambda} \left < m_y^2 \right >_{\Lambda}}\nonumber \\ 
a_y  = & \frac{1}{\gamma_y} \frac{\left < m_y N_{\widehat{\ell}, \overline{\ell}} \right >_{\Lambda} \left < m_x^2 \right >_{\Lambda} - \left < m_x N_{\widehat{\ell}, \overline{\ell}} \right >_{\Lambda} \left < m_y m_x \right >_{\Lambda}}{\left < m_x m_y \right >_{\Lambda}^2 - \left < m_x^2 \right >_{\Lambda} \left < m_y^2 \right >_{\Lambda}} \,
\label{eq:c_x_y}
\end{align}
where the quantities
\begin{align}
m_{x} =  k_{x}^n \widehat{h} \qquad {\rm and} \qquad  \gamma_{x}  = 1 - \left( \frac{\overline{\ell}_x}{\widehat{\ell}_x} \right)^{n-2/3}, \\
m_{y} = k_{y}^n \widehat{h} \qquad {\rm and} \qquad  \gamma_{y}  = 1 - \left( \frac{\overline{\ell}_y}{\widehat{\ell}_y} \right)^{n-2/3}
\end{align}
have been introduced to simplify the notation.  Here, $N_{\widehat{\ell}, \overline{\ell}}  = N[\overline{g}] - N[\widehat{g}] $ represents the sub-test-scale term that is known and can be calculated in a GyroLES simulation.  In addition, the dissipative effect on the model 
is guaranteed by setting to zero any negative coefficient value~\cite{morel13}.
%%%%%%%%%%%%%%%%%%%%%%%%%%%%%%%%%%%%%%%%%%%%%%%%%%%%%%%%%%%%
%%%%%%%%%%%%%%%%%%%%%%%%%%%%%%%%%%%%%%%%%%%%%%%%%%%%%%%%%%%%
%  Numerical results
%%%%%%%%%%%%%%%%%%%%%%%%%%%%%%%%%%%%%%%%%%%%%%%%%%%%%%%%%%%%
\section{Numerical results} \label{sec:numerics}

In the present section, numerical simulations of GK turbulence for different types of instabilities and scenarios, ranging from the well known Cyclone Base Case~\cite{dimits00} to an experimental ASDEX Upgrade discharge, are performed. After introducing   the simulation database for the runs considered, electron density fluctuation spectra will be shown for the different cases.   Finally,   several free energy studies will be presented in the last part of this section.

%%%%%%%%%%%%%%%%%%%%%%%%%%%%%%%%%%%%%%%%%%%%%%%%%%%%%%%%%%%%
%%%%%%%%%%%%%%%%%%%%%%%%%%%%%%%%%%%%%%%%%%%%%%%%%%%%%%%%%%%%
\subsection{Simulation database}

%%%%%%%%%%
\begin{table}[t]
\begin{tabular}{|l|c|c|c|c|c|c|}
   \hline 
   Name  &$\hat{s} $ & $\frac{R}{L_{n}} $&  $\frac{R}{L_{T_i}} $  & $\frac{R}{L_{T_e}} \ $ & grid $(x\times y)$  &  box size $(x\times y)$ \\  \hline
   CBC-H-DNS  &$0.8$ &2.2   & $6.9$ &- &$128 \times 64  $ & $125 \times 125$ \\ 
   CBC-L-DNS &$0.8$ & 2.2  & $6.9$ & -&$64 \times 32    $ & $125 \times 125$ \\
   CBC-LES &$0.8$  &  2.2  & $6.9$ & -&$64 \times 32   $ & $125 \times 125$ \\  \hline
   ITG-H-DNS &$0.8$  & 2.2  & $12.0$ & -&$128 \times 64  $ & $125 \times 125$ \\ 
   ITG-L-DNS&$0.8$   &  2.2 & $12.0$ & -&$64  \times 32    $ &  $125 \times 125$ \\
   ITG-LES &$0.8$  &  2.2  & $12.0$ & -&$64     \times 32    $ & $125 \times 125$ \\  \hline  
   ETG-H-DNS &$0.1$  & 2.2  &- & $6.9$ & $128 \times 128  $ & $200  \times125$  \\ 
   ETG-L-DNS &$0.1$  & 2.2 &- & $6.9$ & $64   \times 64    $ & $200 \times 125$ \\
   ETG-LES &$0.1$   & 2.2 & -& $6.9$ & $64   \times 64    $ & $200 \times 125$  \\ \hline
   TEM-H-DNS &$0.8$  & 3.0  & 0.0& $5.5$ & $128 \times 128  $ & $209  \times104$  \\ 
   TEM-L-DNS &$0.8$  &  3.0 &0.0 & $5.5$ & $64   \times 64    $ & $209 \times 104$  \\
   TEM-LES  &$0.8$  & 3.0 & 0.0& $5.5$ & $64   \times 64    $ & $209 \times 104$  \\ \hline
   AUG-H-DNS&$1.6$   & 0.5  & $5.1$ & $5.1$ & $128 \times 128  $ & $149  \times124$ \\
   AUG-L-DNS &$1.6$  &0.5   & $5.1$ & $5.1$ & $64   \times 64    $ & $149 \times 124$ \\
   AUG-LES &$1.6$  & 0.5  & $5.1$ & $5.1$     & $64   \times 64    $ & $149 \times 124$ \\ 
   AUG-L/2-DNS &$1.6$  &0.5   & $5.1$ & $5.1$ & $32   \times 32    $ & $149 \times 124$ \\
   AUG-LES/2 &$1.6$  & 0.5  & $5.1$ & $5.1$     & $32   \times 32    $ & $149 \times 124$ \\ 
   \hline   
\end{tabular}
\caption{Main parameters for the different simulations.
Horizontal lines separate the relevant sets of data, identified by the same prefix.   
The first set corresponds to the CBC. The second set uses the same parameters as the CBC, but with  a higher  temperature gradient.  The third set  corresponds to a typical  ETG simulation. In this case, a lower magnetic shear  respect to the CBC is used.  The fourth set is used to study  a pure TEM case where  both ions and electrons are kinetic.  Finally, the last set of parameters corresponds to a simulation of an  ASDEX Upgrade discharge  dominated by ITG. 
For every set of parameters, there is a H-DNS ("high resolution" Direct Numerical Simulation), a  L-DNS ( "low resolution" Direct Numerical Simulation) and a LES (Large Eddy Simulation).  The last two cases use  half of the resolution in each of the  perpendicular  directions.  Only for the AUG case,  two extra simulations  with a fourth of the  resolution  are  included. }
\label{tab_database}
\end{table}
%%%%%%%%%%

To analyze the usefulness of  LES methods in  numerical simulations, we look at different cases of GK turbulence, driven by a wide range of instabilities and for different parameter scenarios. As the LES method employed here makes use of a hyper-diffusion model in the $\{x, y\}$ directions, 
we will look at different $\{x, y\}$ resolutions. Meanwhile,  the same $\{16 \times 32 \times 8\}$ resolution is  used in the $\{z \times v_{\parallel} \times \mu\}$ directions for all the cases except the TEM and  AUG simulations which use $\{24 \times 32 \times 16\}$. Velocity space collisional effects are modelled by a fourth-order hyper-diffusivity model in the $z$ and $v_{\parallel}$ directions~\cite{Pueschel}.

Details of the different perpendicular  resolutions and main parameters  for the simulations considered can be found in Table~\ref{tab_database}. The resolutions considered here are used extensively by the fusion community, and for all the cases  the  global transport values (e.g. particle and heat fluxes)  are properly resolved. In general, the lowest resolution  is used as long as these global values are found to vary within $30 \%$. Although in computational fluid dynamics the terminology Direct Numerical Simulations (DNS) is used to denote that all scales are being fully  resolved, we will use DNS in its weak interpretation to denote that no additional sub-grid scale hyper-diffusion model is being used. In this sense, DNS runs should be seen as having varying degrees of incomplete resolution. 
 For this purpose, we will label by  L-DNS the low resolution DNS 
 simulations and by H-DNS the high resolution DNS simulations. The runs for which the sub-grid scale terms are being modelled will be labeled as LES.  

The  first set of parameters corresponds to the Cyclone Base Case, commonly used for the study of  ITG driven GK turbulence, and we label it CBC. In this study, the analysis is limited to the simple scenario of a single ion species and adiabatic electrons in the context of a large aspect-ratio, circular model equilibrium. The equilibrium magnetic configuration is characterized by a safety factor value of $q = 1.4$ and a magnetic shear value of $\hat{s} = 0.8$.

As a way to analyze  the applicability  of LES  methods  for even stronger turbulence regimes, in a second set of parameters we consider  additional simulations with the same parameters as for the standard CBC case, but with a higher  ion temperature gradient ($R / L_{T_i} =12$).
We designate this second set simply as ITG.

The third set is used for the study of a typical ETG driven turbulence, where the adiabatic ion approximation is used~\cite{nevins}. In this case, the LES model acts on the electrons. We will consider again a circular concentric geometry with $q=1.4$ and $\hat{s} = 0.1$. 

The fourth set (designated TEM) is inspired by experiments dominated by electron heating and rather cold ions, specific to turbulence driven by (collisionless) TEMs~\cite{Dannert}. Here, both ion and electron dynamics are retained, which implies that the LES models and their coefficients calculated by the dynamic procedure are   species-dependent. For simplicity, a circular concentric geometry is used with $q=1.4$ and $\hat{s} = 0.8$. In order to study a pure TEM instability,  $R / L_{T_i}$ is set to $0$, and the ratio between the electron and ion temperature is set to $T_e/T_i = 3$ which for these parameters  eliminates  the ETG instability.  It should be noted that such a situation is by no means artificial, since a lot of experiments have been carried out with dominant central electron heating~\cite{Ryter}.

While the above  "idealized" turbulence simulations have the great advantage of minimizing the degree of complexity in performing and analyzing the runs, they usually represent simplified situations which are, in general, of limited value for direct comparisons with experimental findings. For this reason, the last  set applies the dynamic procedure to the study of turbulence for plasma conditions found in
an H-mode ASDEX Upgrade tokamak discharge. The input profile and equilibrium are taken from the ASDEX Upgrade discharge $\#28245$. This discharge is a Type-I edge localized  (ELMy) H-mode with a plasma current of $0.6$~MA and a toroidal magnetic field of $2.3$~T. The input neutral beam injection (NBI) was $2.5$~MW and an electron cyclotron resonance heating (ECRH) was divided into four phases, where  $0.0, 0.5, 1.2$ and $1.8$~MW were applied subsequently at intervals of $0.5$~s. In the following, we will focus on the  phase, where no ECRH is applied, which corresponds to a discharge time of $2.5$~s - $3.0$~s. Furthermore, the local simulations would be focused  on the flux surface at $\rho_{tor}  = 0.57$.   In this case, previous  linear gyrokinetic simulations~\cite{tim13_b}  showed that  ITG is the dominant instability.   ETG  is also present but its relative (to the ITG) amplitude is negligible. For this reason, although we will use both kinetic ions and electrons, we will only resolve  scales in the ITG range.   For this scenario, we use a realistic magnetic equilibrium geometry, taken from the TRACER-EFIT interface~\cite{pablos09}, with equilibrium parameters given as follows: $q=2.8$ and  $\hat{s} = 1.6$. A linearized Landau-Boltzmann collision operator ($\nu_{*,i} = 0.19$  and  $\nu_{*,e} = 0.36$),  the effect of  $E \times B$ shear ($\gamma_E = 0.02 \,[c_s/a]$) and magnetic fluctuations ($\beta = 0.25 \%)$ are included. 
%%%%%%%%%%%%%%%%%%%%%%%%%%%%%%%%%%%%%%%%%%%%%%%%%%%%%%%%%%%%
%%%%%%%%%%%%%%%%%%%%%%%%%%%%%%%%%%%%%%%%%%%%%%%%%%%%%%%%%%%%
\subsection{Electron density fluctuation spectra}

%%%%%%%%%%
\begin{figure*}[p]
\begin{center}
\includegraphics[height = 0.96\textheight]{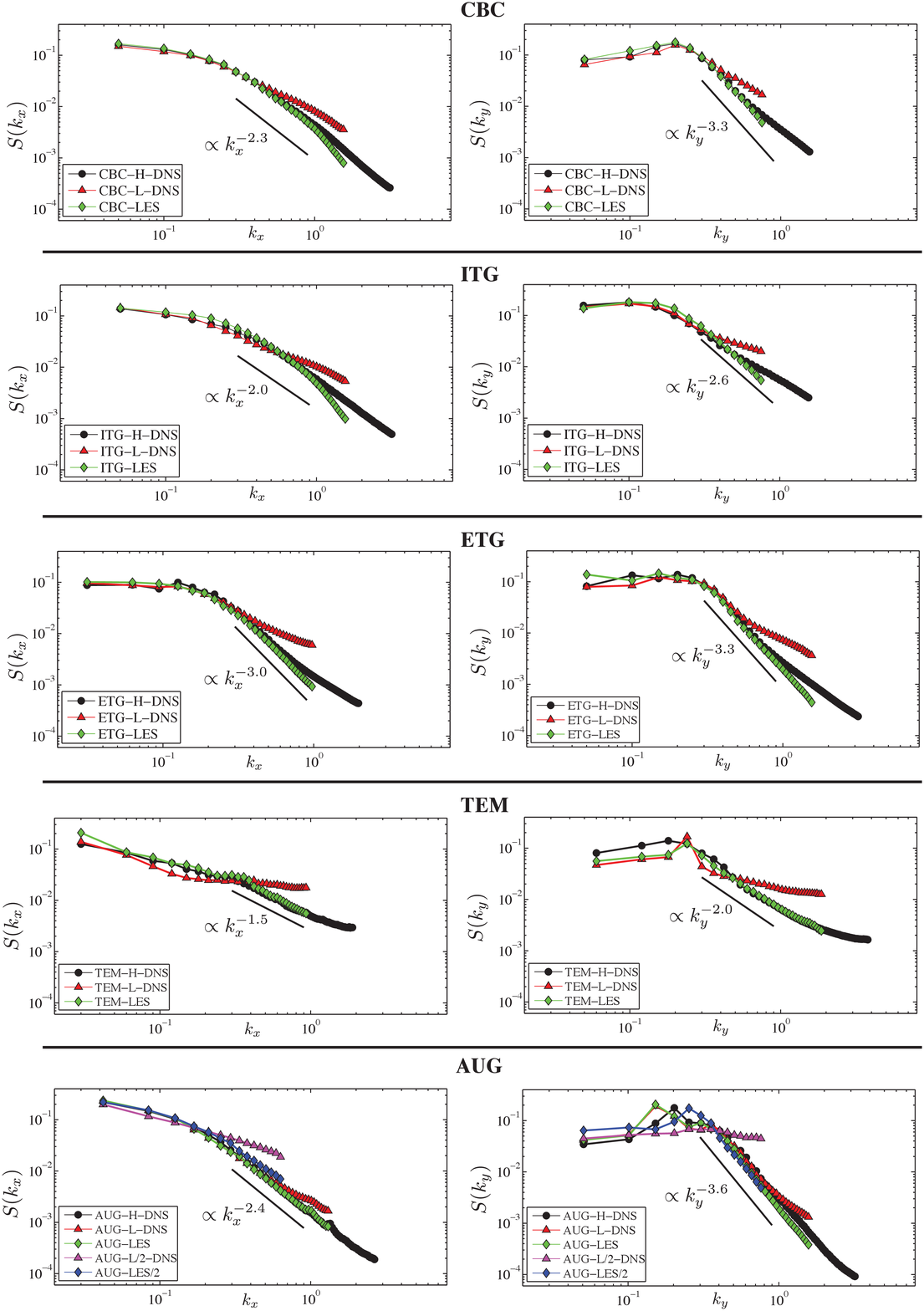}
\end{center}
\caption{(color online) Mean square density fluctuation spectra for the simulations described in Table~\ref{tab_database}. The spectra are normalized by the wavenumber integrated value  and the wavenumbers are in  units of the dominant species gyro-radius ($\rho_i$ for ITG and $\rho_e$ for ETG) for single-species simulations, and in  $\rho_s$ units (ion gyroradius at electron temperature) in the case where two species are considered (TEM and AUG).  }
\label{fig_spec_dens}
\end{figure*}

%%%%%%%%%% 

In the following, the assessment   of LES methods compared to various DNS runs of different resolutions are shown for  the electron density fluctuation spectra.
In particular,   the electron density fluctuation spectrum in the binormal direction $S(k_y) = \left<  | \tilde{n}_{e} (k_x, k_y, z, t) |^2  \right >_{k_x, z, t} $  as well  as in the radial direction $S(k_x) = \left<  | \tilde{n}_{e} (k_x, k_y, z, t) |^2  \right >_{k_y, z, t} $  are plotted in Fig.~\ref{fig_spec_dens}.  Here,  $\left <  \cdots  \right >$ denotes  averaging over quantities listed as indices  and  all spectra are  normalized by their respective wavenumber integrated value $ \left<  | \tilde{n}_{e} (k_x, k_y, z, t) |^2  \right >_{k_x, k_y, z, t}$. The  wavenumbers are normalized in  units of the dominant species gyroradius ($\rho_i$ for ITG and $\rho_e$ for ETG) for one kinetic species simulations, and in  $\rho_s$ units (ion gyroradius at electron temperature) in the case where two kinetic species are considered (TEM and AUG). Some general features are common to  all data sets: the $S(k_y)$ spectra   exhibit  a  maximum at $k_y  \sim 0.1- 0.2$   and the radial $S(k_x)$ spectra peak at wavenumbers close to zero.  In both cases,  a power law  $S(k_{x,y}) \propto k_{x,y}^{-\alpha_{x,y}}$ for  wavenumbers between $k_{x,y} \sim 0.3 - 1.0$ is observed.  Although, a transition and a change of the power law is expected by several theories at  $k_{x,y}   >  1$ (see Ref.~[\onlinecite{plunk10}]), we will limit our study   up to $k_{x,y}   \sim 1$. Therefore, in the following comparisons of the spectra will be focus on   the   wavenumber range  between $k_y  \sim 0.3 - 0.9$, where a fit to the power law exponents is given for the LES simulations.  

For  the  CBC set of parameters, a fit of spectra yields the    power law exponents    $\alpha_x = 2.3$ and $\alpha_y = 3.3$. In this case, the  LES (in green)  spectra match very well the H-DNS (in black) spectra in both $S(k_x)$ and $S(k_y)$. In contrast,    L-DNS (in red)  spectra get  flattened at higher wavenumbers. For the higher temperature  gradient  ITG case, the fit  exponents  are $\alpha_x = 2.0$ and $\alpha_y=2.6$. Regarding  the $S(k_x)$ spectra,  similar  conclusions as in the CBC  can be drawn.  However, the $S(k_y)$ spectrum  exhibits   a bigger difference between the LES and  H-DNS simulations. In fact, the H-DNS spectra seem to present a flattening of the spectra at the highest wavenumbers.  
We  anticipate now (more details are given in the next section)  that this  is due to an accumulation of free energy. Indeed,  since this case  represents a  stronger turbulent  case compared to the CBC, the importance of numerically   removing   accurately the energy at smaller scales  becomes more  important. The ETG set of parameters  is another  example of the  flattening of  spectra even at  the H-DNS resolution, observed in this case in both spectra.
These simulations, dominated by streamers, can be considered   as an equivalent  of  
 stronger turbulent simulations. For this case, the exponents  are  $\alpha_x = 3.0$ and $\alpha_y = 3.3$.  The fourth set of simulations, given by TEM turbulence, power law exponents  of   $\alpha_x = 1.5$ and $\alpha_y = 2.0$ are found. Here,  a good agreement between  the  H-DNS  and LES simulations is observed. In contrast, the L-DNS presents now practically  flat spectra.  Therefore, it seems evident that for this case  the  use of LES methods  for   the low resolution simulations is needed. 

The previous cases  show how the LES procedure can  be successfully applied  to different types of microturbulence. However, these  are simple setups, and the resulting  exponents cannot be compared directly with experimental measurements.   For this reason, we finally also consider a realistic example of ITG turbulence. Here, the power law exponents are $\alpha_x = 2.4$ and $\alpha_y=3.6$.   Interestingly, there is a good agreement between AUG-H-DNS, AUG-LES,  and AUG-L-DNS, although the latter  displays  a flattening of the spectra at the highest wavenumbers. Moreover, it is possible to further decrease the resolution without changing the values of  the heat and particle fluxes. For this reason, two additional  simulations are included,  AUG-L/2-DNS (in purple)  and AUG-LES/2  (in blue), see Fig~\ref{fig_spec_dens}. Now the differences are  more evident: while  AUG-LES/2 overlaps perfectly with the LES and H-DNS simulations,  L/2-DNS exhibits  flat spectra. This shows that for some cases with very  limited resolution,  LES methods can succeed in recovering the correct  power law exponents for experimentally relevant  cases.
 
\begin{figure}[!hbt]
\begin{center}
\includegraphics[width= 1.0\columnwidth]{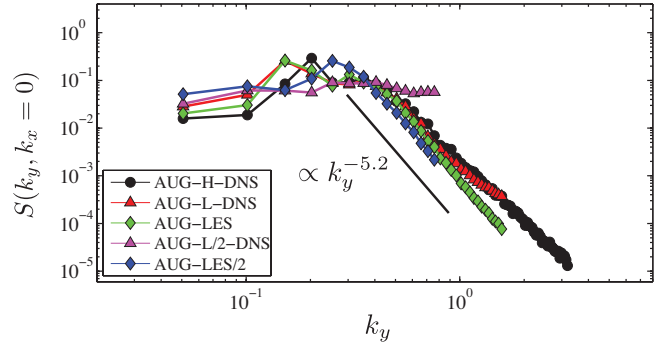}
\end{center}
\caption{(color online). Mean square density fluctuation spectra $S(k_y, k_x=0)$    in the outboard mid-plane for the AUG data set. The spectra are normalized by the wavenumber integrated value, and the wavenumbers are in $\rho_s$ units.}
\label{fig_spec_dens_kx0}
\end{figure}

At this point, it is  worth to mention the anisotropy observed in the simulations for all the cases, see Table~\ref{tab_density} for a summary of the results. In general, the $\alpha_{y}$ exponents are higher than the $\alpha_{x}$. Such deviations from isotropy  should be taken into account  when comparing numerical with experimental results. In particular,  because in the experimental measurements often consider $k_x = 0$ contributions and  the measurements are done   in the outboard mid-plane ($z=0$ plane in {\sc gene}). 
Therefore,  for the AUG dataset   the  $S(k_y, k_x = 0) = \left<  | \tilde{n}_{e} (k_x, k_y, z, t) |^2  \right >_{k_x=0,z=0,t}$  spectrum is shown in  Fig.~\ref{fig_spec_dens_kx0}. In this case, the calculated exponent  is higher,  rising to  a value of $\alpha_y = 5.2$. Moreover, the recovery of the same $\alpha_y = 5.2$ value by the two LES runs possessing different resolutions, shows the tendency of LES methods to converge on the correct dynamical results.
This behavior is not found by the DNS runs (L and L/2 runs differ drastically from each other).

Finally, in addition to the shape of the spectra, it is also important to calculate the wavenumber integrated value
$\left<  \tilde{n}^2(k_x,k_y,z) \right>_{k_x,k_y,z}$.  
However, since small scales are truncated in the LES and  L-DNS runs, it is not possible to integrate up the same scale as in the H-DNS simulation.  Therefore, an estimate of the contribution from the truncated scales is certainly desirable if a comparison has to be made with experimental results. For this reason,  an estimate for the truncated scale contribution is proposed by fitting a power-law to the spectrum and extrapolating to unresolved scales when integrating to the the total fluctuation level. (For specific  details see  Ref.~\onlinecite{morel11}.)  The results  are summarized in  Table~\ref{tab_density}. While the differences regarding the wavenumber integrated  electron density  for LES simulations can exceed $45 \%$ for the  lowest resolution AUG case,  simulations without a LES model are even much more inaccurate, exhibiting relative errors up to about $200 \%$.

To summarize, we have shown that  LES methods  provide 
a better accuracy in the calculation of   power law exponents   for different scenarios and type of instabilities. Since the  use of LES does not increase the cost of the simulations in comparison with normal (DNS) simulations with the same resolution, it should be considered whenever possible. In particular, LES   behaves   better than  simulations  with two times  more  resolution in each of the  perpendicular coordinates,   at  fraction of the cost (at least $4$ times cheaper than the H-DNS simulations).
\begin{table}[!h]
\begin{tabular}{|c|c|c|c|c|c|}
   \hline 
    Name            &CBC      &ITG     &ETG       &TEM     &AUG   \\ \hline
    $\alpha_x$   & $2.3$    &$2.0$   &$3.0$     &$1.5$   &$2.4$ \\ \hline
    $\alpha_y$   &$3.0$     & $2.2$  & $3.6$    &$2.0$   &$3.2(5.2)$ \\ \hline
    $\left< \tilde{n}^2\right >^{\rm \tiny{LES}} /  \left< \tilde{n}^2\right >^{\rm \tiny{DNS}}$ & $0.87$    &$0.98$   &$1.26$     &$1.41$   &$0.71$ \\ \hline
    $\left< \tilde{n}^2\right >^{\rm \tiny{L-DNS}} /  \left< \tilde{n}^2\right >^{\rm \tiny{DNS}}$ & $1.02$    &$1.79$   &$0.95$     &$2.11$   &$0.63$ \\ \hline
    $\left< \tilde{n}^2\right >^{\rm \tiny{LES/2}} /  \left< \tilde{n}^2\right >^{\rm \tiny{DNS}}$ & $-$    &$-$   &$-$     &$-$   &$0.52$ \\ \hline
    $\left< \tilde{n}^2\right >^{\rm \tiny{L/2-DNS}} /  \left< \tilde{n}^2\right >^{\rm \tiny{DNS}}$ & $-$    &$-$   &$-$     &$-$   &$1.3$ \\ 
    \hline    
\end{tabular}
\caption{Fitted power law exponents   for the  density fluctuation spectra, together with the wavenumber  integrated value  of  the  electron  density  for the LES and L-DNS simulations normalized to the total   value  of the  H-DNS simulation.
For the AUG case, the value in brackets indicates the power law exponent for the 
$S(k_y, k_x =0)$ spectrum.}
\label{tab_density}
\end{table}
%

%%%%%%%%%%%%%%%%%%%%%%%%%%%%%%%%%%%%%%%%%%%%%%%%%%%%%%%%%%%%
%%%%%%%%%%%%%%%%%%%%%%%%%%%%%%%%%%%%%%%%%%%%%
\subsection{Free energy  studies}
\begin{figure*}[p]
\begin{center}
\includegraphics[height = 0.96\textheight]{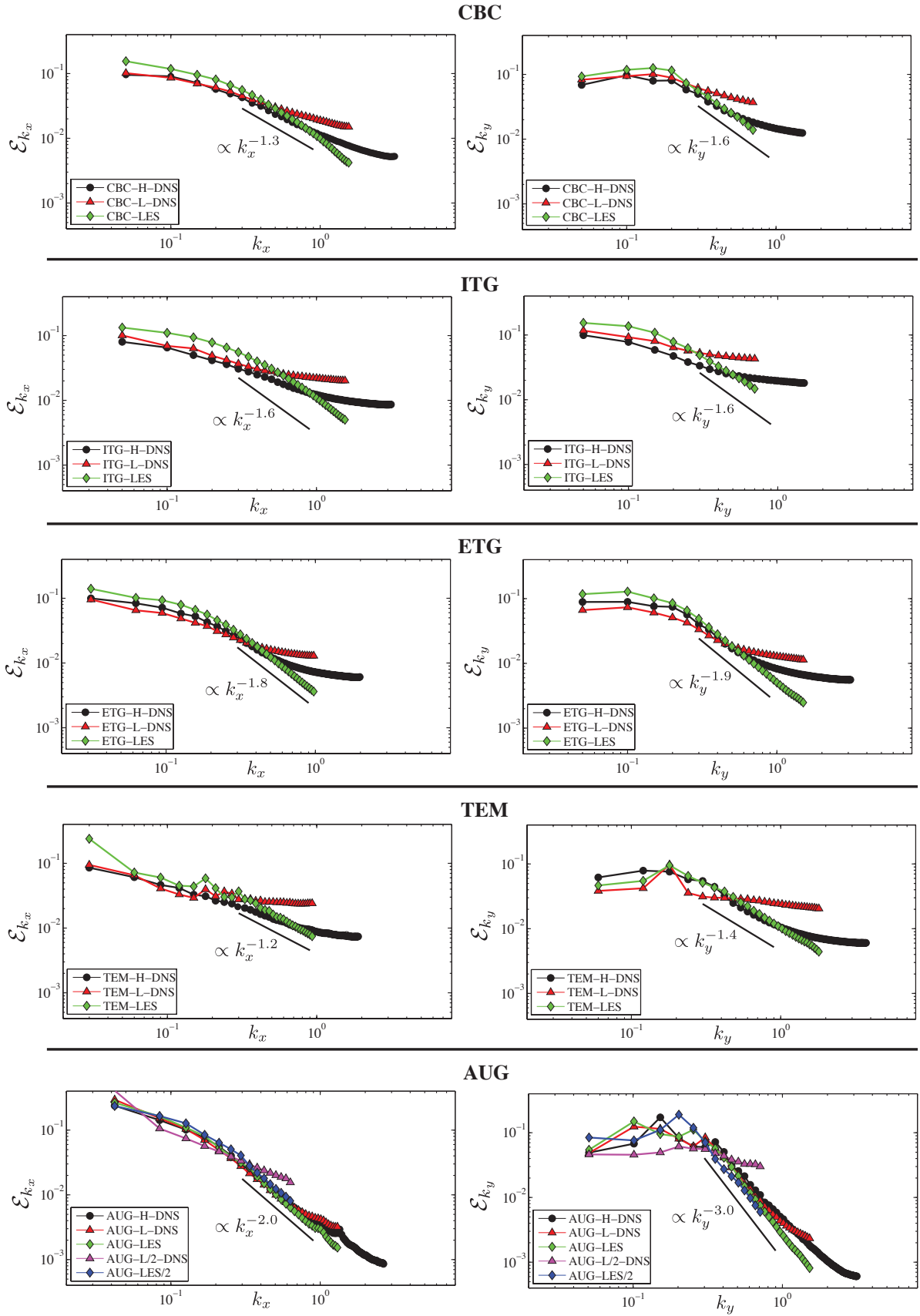}
\end{center}
\caption{(color online) Free energy spectra for all the simulations described in Table~\ref{tab_database}. The spectra are normalized by the wavenumber integrated value  and the wavenumbers are in  units of the dominant species gyro-radius ($\rho_i$ for ITG and $\rho_e$ for ETG) for single-species simulations and in  $\rho_s$ units (ion gyroradius at electron temperature) in the case where two species are considered (TEM and AUG).}
\label{fig_spec_ener}
\end{figure*}

The previous analysis looked  at the density fluctuation spectra. However, although not measurable experimentally, it is the free energy (see Ref.~\onlinecite{scheko09}) which  determines the resulting power laws observed in other quantities (such as density/temperature fluctuations).
The study of the free energy is also important to understand the dynamics of the system and the range of validity of LES methods.   The free energy ($\mathcal{E}=\mathcal{E}^f + \mathcal{E}^{ \phi} + \mathcal{E}^{A}$), consisting in the mixing of entropy ($\mathcal{E}^f$), electric ($\mathcal{E}^{ \phi}$) and ($\mathcal{E}^{A}$) magnetic energies, is the quantity that is injected into the system by the gradients and dissipated by collisions. Moreover, free energy is redistributed between different scales by the action of the nonlinear term, without global gains or losses. The global free energy is also known as a nonlinear invariant quantity and has been proved to have many similarities with the kinetic energy in fluid turbulence~\cite{banon11}. For these reasons, in the following  we will introduce and analyze in  detail different free energy diagnostics.

%%%%%%%%%%%%%%%%%%%%%%%%%%%%%%%%%%%%%%%%%%%%%%%%%%%%%
\subsubsection{Free energy  fluctuation spectra}

Formally, the free energy spectral density is defined as
\begin{align}
\mathcal{E}_k  = \left<g_{k} \left[\frac{n_{0} T_{0}}{2\, F_{0}}h_k \right]\right>_{z,v_\parallel, \mu ,j}\,,
\end{align}
where  $\left<  \cdots  \right>$ represents an integration over the listed index.
The background density and temperature level  is given by $n_{0}$ and $T_{0}$, respectively.   $F_{0}$ represents the Maxwellian contribution to the total distribution function. 

The free energy fluctuation spectra  are plotted in Fig.~\ref{fig_spec_ener} with regard to $k_x$ and $k_y$.  All spectra   ($\mathcal{E}_y$ and $\mathcal{E}_x$) are  normalized
by their respective wavenumber integrated value ($ \mathcal{E} = \left < \mathcal{E}  \right>_{k_x,k_y}$). As before,  the free energy spectrum in the binormal direction $\mathcal{E}_y$  peaks at $k_y \propto  0.1-0.2$ and a power law $\mathcal{E}_y \propto k_y^{\beta_y}$  is present. The radial wavenumber spectrum $\mathcal{E}_x$ peaks at $k_x = 0$ and has also  a power law $\mathcal{E}_x \propto k_x^{\beta_x}$ for higher wavenumbers.  Table~\ref{tab_free_energy} shows a summary with the power law exponents calculated for the different data sets. For all cases,   the anisotropy in the spectra  is also found.  In addition, the  wavenumber integrated value of the free energy is   shown in Table~\ref{tab_free_energy}. Comparing to the H-DNS value,
the total free energy can exceed  $250 \%$ for L-DNS simulations,  while for the LES simulations   
differences  only  up  to $40 \%$ are found.
 
Analyzing in   detail  the different cases,    similar conclusions as  for the  density fluctuation spectra can be drawn for all cases, although now  the differences  between H-DNS, L-DNS and LES simulations are  more evident: LES simulations present always a power law, while both H-DNS and L-DNS presents a more clear  flattening of the spectra. 
For instance, looking at the   ITG data set, and in particular  at the   H-DNS simulation, we observe 
that it  has a  flat spectrum.  This  is also the case for the L-DNS simulation. However, the LES still  presents    a  power law.  The reason for this behavior is the free energy accumulation of the DNS runs, which again shows that DNS runs are in fact, to different degrees, unresolved simulations. For a proper DNS run, the  tail of the spectrum is expected to decrease in value at a faster rate than in the cascade range and not to posses a shallower slope. 

Thus,  the free energy is a good indicator to check if a simulation is well resolved. Looking at the spectra, and in particular the small scales behavior, one can distinguish if a simulation requires a larger resolution or not. An accumulation of energy at high wavenumbers will quickly amount to a change in the nonlinear dynamics and is undesirable. As the LES models are derived from  the nonlinear transfers, the energy can be seen as being transferred to the unresolved range of scales  rather than being arbitrarily removed. These effects, although being more evident for the  free energy, as discussed in the previous section,  are also present in other relevant quantities, such as potential,  density  or temperature fluctuations.  Finally, considering the spectral slope extension into the unresolved range of wavenumbers, it can be seen than a LES run can provide better results than a high resolution DNS at a fraction of the cost.   
\begin{table}[!h]
\begin{tabular}{|l|c|c|c|c|c|}
   \hline 
    Name            &CBC      &ITG     &ETG       &TEM     &AUG   \\ \hline
    $\beta_x$   & $1.3$    &$1.6$   &$1.8$     &$1.2$   &$2.0$ \\ \hline
    $\beta_y$   &$1.6$     & $1.6$  & $1.9$    &$2.0$   &$3.0 $ \\  \hline
    ${\mathcal{E}}^{\rm \tiny {LES}}/ {\mathcal{E}}^{\rm \tiny {DNS}}$  &$0.84$     & $0.75$  & $1.05$    &$1.39$   &$0.85 $ \\ \hline
    ${\mathcal{E}}^{\rm \tiny {L-DNS}}/{\mathcal{E}}^{\rm \tiny {DNS}}$  &$1.71$     & $2.76$  & $2.29$    &$2.12$   &$0.82 $ \\ \hline
    ${\mathcal{E}}^{\rm \tiny {LES/2}}/{\mathcal{E}}^{\rm \tiny {DNS}}$  &$-$     & $-$  & $-$    &$-$   &$0.75 $ \\ \hline
    ${\mathcal{E}}^{\rm \tiny {L/2-DNS}}/{\mathcal{E}}^{\rm \tiny {DNS}}$  &$-$     & $-$  & $-$    &$-$   &$1.65$ \\ 
    \hline    
\end{tabular}\caption{Fitted power law exponents  for the free energy  fluctuation spectra, together with the wavenumber  integrated value  of  the   free energy    for the LES and L-DNS simulations normalized to the total   value  of the  H-DNS simulation.}
\label{tab_free_energy}
\end{table}
%

%%%%%%%%%%%%%%%%%%%%%%%%%%%%%%%%%%%%%%%%%%%%%%%%%
\subsubsection{Nonlinear transfer spectra}

Until now, we have discussed the free energy spectra. In order to understand how these spectra are formed, we need to study the nonlinear cross-scale transfer of free energy. The corresponding spectral balance equation has the form
\begin{align}
\frac{\partial  \mathcal{E}_k } {\partial t}  = \mathcal{L}_{k}  + \mathcal{D}_k+  \mathcal{T}_k\;.
\label{eq:FE_spectral}
\end{align}
Here, $\mathcal{L}_k$ represents the linear contributions composed by the free energy injected into the system at scale $k$ (by the temperature/density gradients) as well as the contributions for the parallel and curvature terms. The term  $\mathcal{D}_k$ is the local dissipation, and $\mathcal{T}_k$ is the nonlinear free energy transfer term. The latter represents the redistribution of free energy between all modes $\mathbf  k$ that contribute to a scale $k$, due to the interaction with modes $\mathbf  p$ and $\mathbf  q = -\mathbf k - \mathbf p$, i.e. all triad interactions that have the $k$ scale as one of the legs. Formally, the interaction between three scales can be defined as
\begin{align}
\mathcal T_{k| p, q}\!= \int_{|\mathbf k|=k} \!\!\!\!\! \mbox{d}\mathbf k \int_{|\mathbf p|=p}\!\!\!\!\! \mbox{d}\mathbf p \int_{|\mathbf q|=q} \!\!\!\!\!\mbox{d}\mathbf q\;
 \mathcal T_{\mathbf k| \mathbf p, \mathbf q}\; \delta({\mathbf k}\!+\!{\mathbf p}\!+\!{\mathbf q})\;, 
\label{trip}
\end{align}
where the fundamental triad transfer has the form
\begin{align}
\mathcal T_{\mathbf k| \mathbf p, \mathbf q}\!= \left < \frac{n_0 T_{0}}{2F_{0}} \Big{[} q_x p_y \!-\!q_y  p_x \Big{]}\!\! \Big{[} \chi_{\mathbf q} h_{\mathbf p} \!-\!\chi_{\mathbf p} h_{\mathbf q} \Big{]}  h_{\mathbf k} \right >_{z,v_\parallel,\mu,j} \;. 
\label{trip}
\end{align}

For the GyroLES approach, it is also possible to write the spectral free energy balance equation for resolved scales,
\begin{align}
\frac{\partial} {\partial t}  \mathcal{E}^{<}_{k | k_c}   = \mathcal{L}^{<}_{k | k_c} +  \mathcal{D}^{<}_{k | k_c} + \mathcal{T}^{<}_{k | k_c} +  \mathcal{T}^{SGS}_{k | k_c}. 
\label{eq:FE_spectral_LES}
\end{align}
The  sub-grid transfer $\mathcal{T}^{SGS}_{k | k_c}$ represents  the transfer of energy between resolved scales $k < k_c$  and sub-grid scales  $k > k_c$. It is related to the  free energy transfer by
\begin{align}
\mathcal{T}^{SGS}_{k | k_c} = \mathcal{T}_k  -  \mathcal{T}^{<}_{k | k_c}.
\label{eq:Tsplit}
\end{align}
This equation provides a simple method to compute $\mathcal{T}^{SGS}_{k | k_c}$ through two calculations of the transfer terms. It consists of taking a Direct Numerical Simulation (DNS) and a test-filter DNS simulation  at the characteristic scale $k_c$. Since all the information of free energy transfer $\mathcal{T}_k$ is a available in addition to the largest scale one, we can also calculate  $ \mathcal{T}^{<}_{k | k_c}$ as the difference of the two. This method was used in previous works~\cite{morel11,morel13} to study the properties of the sub-grid transfer. It was found out that its effect is to systematically dissipate free energy from the system. This is the main reason behind choosing a hyper-diffusion LES model, as it can be proved analytically that for positive free parameters this term dissipates free energy at all times. However, in the dynamic procedure introduced in the previous section, there is not a constraint regarding the sign of the free parameter, and in fact, sometimes it can be negative. For this reason, the free energy dissipative effect of the model is guaranteed by setting to zero any negative coefficient values in Eq.~\eqref{eq:c_x_y}.

\begin{figure}[!htb]
\begin{center}
\includegraphics[width= 1.0\columnwidth]{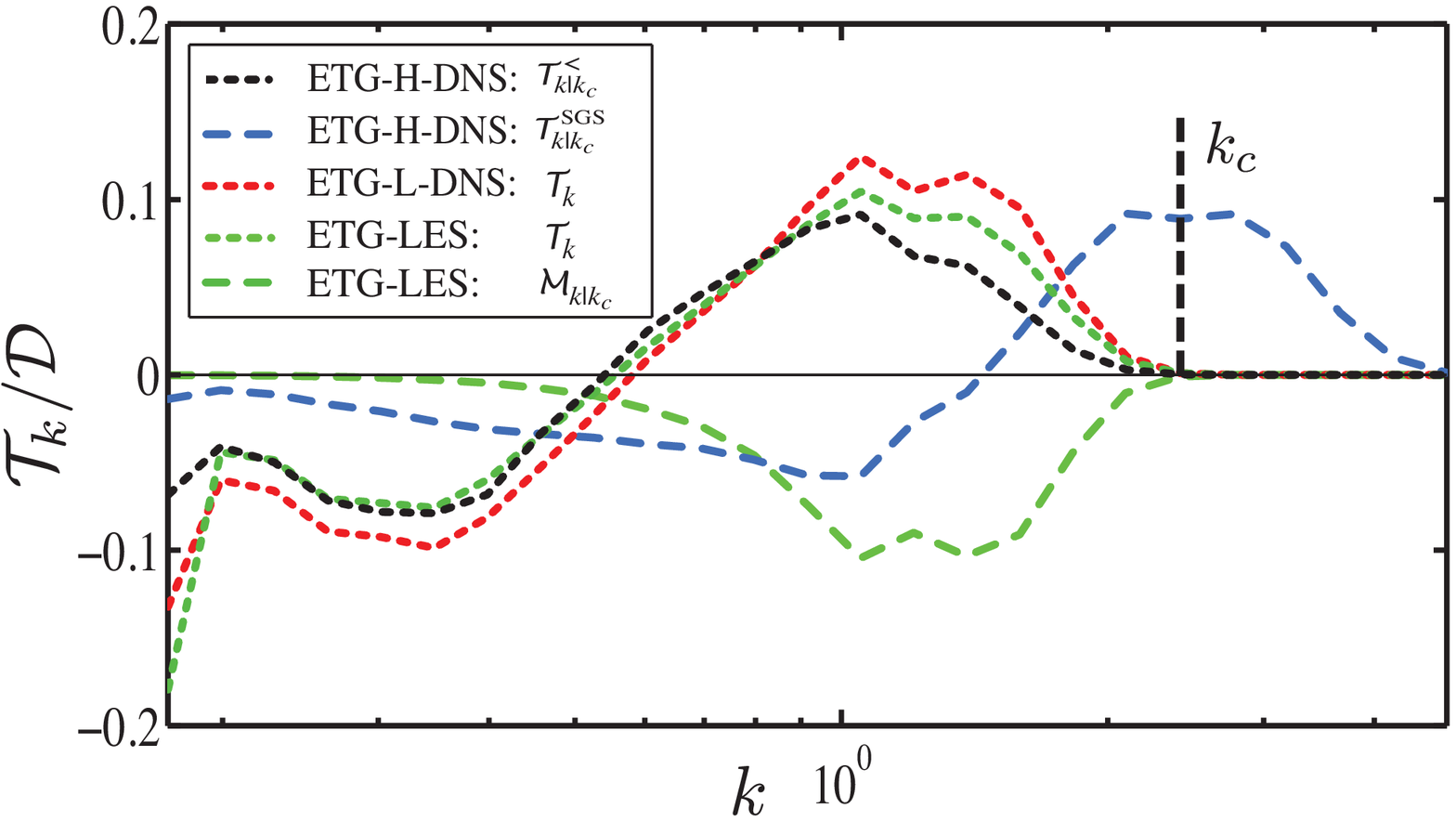}

\end{center}
\caption{Nonlinear free energy transfer spectra for the ETG set of data. The vertical dotted lines denote maximal scales available for the  smaller resolutions.}
\label{fig_tra_spec}
\end{figure}

In the following, we will study the free energy transfers defined in terms of the perpendicular wavenumber  $k=[g^{xx}k^2_x+2g^{xy}k_xk_y+g^{yy}k^2_y]^{1/2}$ which is directly related to physical scales (in contrast to $k_x$ and $k_y$). Here, $g^{xx}$, $g^{xy}$ and $g^{yy}$  the  metric coefficients associated with the field-aligned coordinate system~\cite{lapillonne09}. In the free energy balance equation, Eq.~\ref{eq:FE_spectral}, the nonlinear free energy transfer ($\mathcal T_k$) represents the energy received by a scale ($k$) from the interaction with all other scales in the system. A positive value indicates that energy is received, while a negative one shows that energy is in fact removed from that scale. Unlike linear quantities, reducing the resolution available to the system limits the interactions between scales and changes the $\mathcal T_k$ spectra. To see this effect and the implication on LES methods,  we concentrate on the ETG data set, although similar conclusions can be obtained for the other cases.   In Fig.\ref{fig_tra_spec} we plot the  spectral decomposition of the transfer $\mathcal T_k$  into the transfer $\mathcal T^<_{k|k_c}$ (dotted-black line)  arising from the interaction of solely large scales ($k<k_c$) and the transfer spectra $\mathcal T^{SGS}_{k|k_c}$  (dashed-blue line) involving all other interactions. Since $k_c$ is the maximal scale obtained by halving the ETG-H-DNS resolution, it is clear that a large portion of computation costs is dedicated to a small dynamical range. However, this small range cannot be simply removed, as its effect on $\mathcal T_k$ is evident, i.e., $\mathcal T^<_{k|k_c}$ and $\mathcal T^{SGS}_{k|k_c}$ are comparable in amplitude. 

For a LES run,  the $\mathcal T^<_{k|k_c}$ signal is computed directly ($\mathcal{T}^<_{k|k_c} = \mathcal{T}_k$) while the $\mathcal T^{SGS}_{k|k_c}$ is accounted by the model. The model contribution (dashed-green line)  and the actual $\mathcal T^{SGS}_{k|k_c}$ signal (dashed-blue line) are in the same order of magnitude for low $k$, but start to deviate when $\mathcal T^{SGS}_{k|k_c}$ changes its     character from a sink to a source. This is to be expected, as the model amplitude obtained from the dynamical procedure is always taken to be positive for a hyper-diffusivity LES model. Looking at the resolved transfer spectra ($\mathcal T^<_{k|k_c}$, dotted-black line) we see a good agreement with the LES transfer spectra ($\mathcal{T}_{k}$, dotted-green line). Moreover, considering the ETG-L-DNS run, which has the same resolution as ETG-LES and differs from the ETG-H-DNS runs by the $\mathcal T^{SGS}_{k|k_c}$ term, we see  that the  low  resolution DNS transfer spectra ($\mathcal{T}_{k}$, dotted-red line)  deviates more form the resolved spectra than the LES run.

\begin{figure}[!htb]
\begin{center}
\includegraphics[width= 1.0\columnwidth]{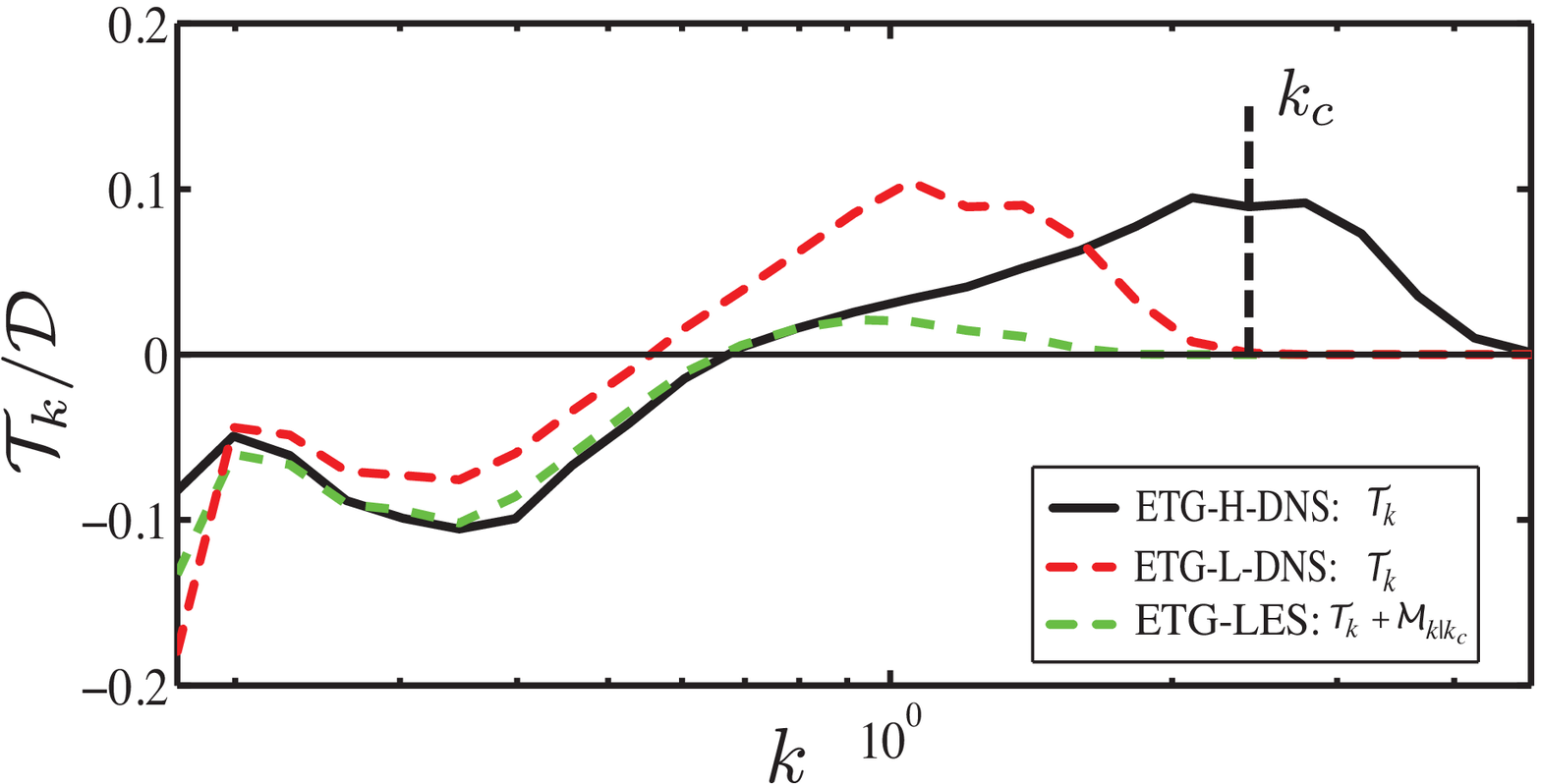}

\end{center}
\caption{Nonlinear free energy transfer spectra for the ETG set of data. The vertical dotted lines denote maximal scales available for the  smaller resolutions.}
\label{fig_tra_les}
\end{figure}

To account for the large resolution DNS (ETG-C-DNS) free energy transfer ($\mathcal T_k$), both the LES resolved and the model contributions need to be considered. In Fig.~\ref{fig_tra_les} we  plot the sum of these contributions to a  LES run.    We can see that smaller DNS runs (obtained in the absence of a model) generate a transfer spectra that deviate more and more compared to the largest DNS one at low $k$. In comparison, the LES transfer spectra  plus  the model contribution try to match the DNS transfer curve, partially successful at lower $k$, regardless of the cutoff. 

%%%%%%%%%%
%%%%%%%%%%

%%%%%%%%%%
%%%%%%%%%%

%%%%%%%%%%%%%%%%%%%%%%%%%%%%%%%%%%%%%%%%%%%%%
\subsubsection{Shell-to-shell transfer}

The shell-to-shell transfer represents an additional diagnostic that can show the advantage of the LES method. The diagnostic consists in filtering the the distribution function and  considering only the modes contained in shell like structures $s_K=[k_{K-1},k_{K}]$, before building the free energy transfer functions. The boundary wavenumbers ($k_{K}$) are given as a geometric progression, here $k_K=k_0 \times 2^{(K-1)/5}$, and the shell-filtered distribution functions $g^K_{\mathbf k}$ are given by
\begin{eqnarray}
g^K_{\mathbf k}=\left\{ \begin{array}{lcl}
g_{\mathbf k} , &  |{\mathbf k}| \in s_K \\
0 , &  |{\mathbf k}| \notin s_K
\end{array}  \right.. 
\end{eqnarray}
It is important to realize that the shell-filtered distribution functions are well defined in real space, the total signal being recovered as the superposition of all scale filtered contributions, $g=\sum_K g^K$. As the time evolution of a shell-filter signal due solely to the nonlinear term  can be expressed as
\begin{align}
\frac{\partial g^K}{\partial t}\bigg{|}_{N}= \sum_{P}\ \Big{[}\frac{\partial \chi}{\partial y}\frac{\partial h^P}{\partial x} - \frac{\partial \chi}{\partial x}\frac{\partial h^P}{\partial y}\Big{]}\;,
\label{nonlin}
\end{align}
the resulting spectral free energy triad-transfers have the form
\begin{align}
\bar{ \mathcal T}_{\mathbf k| \mathbf p, \mathbf q}\!=\left < \frac{n_0 T_{0}}{2F_{0}} \Big{[} q_x p_y \!-\!q_y  p_x \Big{]}\!\! \Big{[} \chi_{\mathbf q} h^P_{\mathbf p} \!-\!\chi_{\mathbf p} h^P_{\mathbf q} \Big{]}  h^K_{\mathbf k} \right>_{z,v_{\parallel},\mu,j} \;. 
\label{trip}
\end{align}
For $\bar{ \mathcal T}_{\mathbf k| \mathbf p, \mathbf q}$, the manifest symmetry in ${\mathbf q}$ and ${\mathbf p}$ of the triad transfers is broken effectively by the shell filtering procedure, as $h^P_{\mathbf q}=0$ for $q\notin s_P$. 

The shell-to-shell transfer is then defined simply as
\begin{align}
\mathcal P^{K,P}\!= \!\int_{|\mathbf {k}|\in s_K}\! \!\!\!\!\! \mbox{d} {\mathbf k} \int_{|\mathbf {p}|_\in s_P} \!\!\!\!\!\! \mbox{d}{\mathbf p} \int_{-\infty}^{\infty}\!\!\!\!\ \mbox{d}{\mathbf q}\, \bar{ \mathcal T}_{\mathbf k| \mathbf p, \mathbf q}\; \delta({\mathbf k}\!+\!{\mathbf p}\!+\!{\mathbf q})\;.
\label{s2s}
\end{align}
It has the interpretation of the energy received by modes located in a shell $K$ from modes located in a shell $P$ by the interaction with all other possible modes. Due to the conservation of interaction, $\mathcal P^{K,P}=-\mathcal P^{P,K}$ and $\mathcal P^{K,K}=0$ for each species. Since the shell boundaries are taken as a power law, the normalized results to the maximal shell transfer, provides us with information regarding the direction and locality of the energy cascade. We designate a transfer to be direct if it is positive for $K>P$ and we call it to be local if $|K-P|\sim5$. 

\begin{figure}[!htb]
\begin{center}
\includegraphics[width= 1.0\columnwidth]{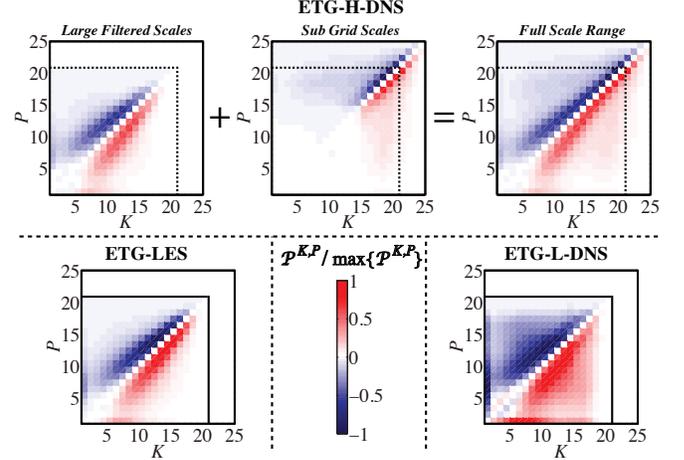}
\end{center}
\caption{Shell-to-shell transfer for the ETG case. The dotted line plotted for the H-DNS run corresponds to the LES wavenumber filter boundary. All transfers for a run are normalized to the respective run maximal value and $k_0=0.173$ in the shell boundary power law.}
\label{fig_S2S_ETG}
\end{figure}

\begin{figure}[!htb]
\begin{center}
\includegraphics[width= 1.0\columnwidth]{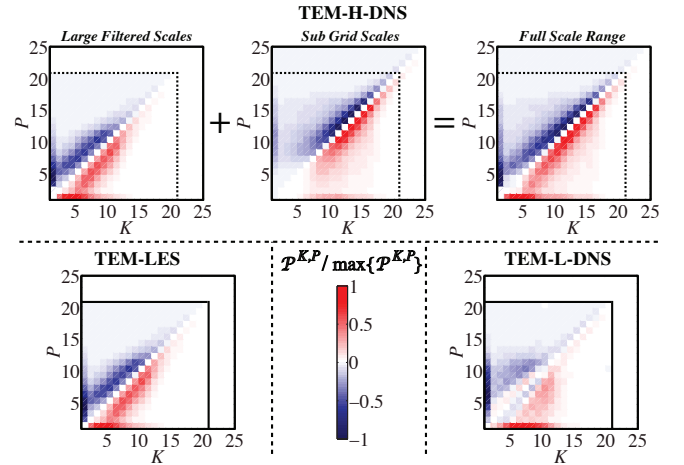}
\end{center}
\caption{Shell-to-shell transfer for the TEM case, summed over all species. The dotted line plotted for the H-DNS run corresponds to the LES wavenumber filter boundary. All transfers for a run are normalized to the respective run maximal value and $k_0=0.439$ in the shell boundary power law.}
\label{fig_S2S_TEM}
\end{figure}

In Fig.~\ref{fig_S2S_ETG} we look at the shell-to-shell transfer for the ETG case. The dotted line plotted for the ETG-H-DNS run represents the $k$ boundary induced by the LES wavenumber filter. It is interesting to note that while the resolved scales shell transfers (obtained implicitly from Eq.~\eqref{trip} by applying the LES filter before the shell filters) are clearly bounded by this limit, the SGS shell-to-shell transfers penetrate strongly below it, indicating that wavenumbers larger than the LES cutoff contribute to scales smaller than $k_c$. From this picture, the advantage of the LES method is obvious. The cascade recovered by the LES run behaves in a good part as the large filtered scales cascade for the larger DNS run. In comparison, the reduced DNS run (ETG-L-DNS) has stronger off-diagonal contributions and even exhibits a change in the direction of the cascade for the first few shells. This change in the direction of the cascade, for a limited resolution DNS, can also be seen in the case of TEM, see Fig.~\ref{fig_S2S_TEM}. However, while this is strong effect for TEM, almost no effect is observed for ITG driven simulations. For the CBC, ITG and AUG cases, the small resolution DNS runs have a very similar form compared to their respective large resolution counterparts. 

%%%%%%%%%%
%%%%%%%%%%

%%%%%%%%%%
%%%%%%%%%%

%%%%%%%%%%%%%%%%%%%%%%%%%%%%%%%%%%%%%
\subsubsection{Free energy fluxes}

In addition to the spectral free energy balance equation for a scale $k$, it is also worth to look at the free energy contained by all the scales larger than $k$. Integrating the resolved scales free energy balance equation \eqref{eq:FE_spectral_LES}, we find,
\begin{align}
\frac{\partial} {\partial t} \int_{0}^{k} \! \!\mathcal{E}^{<}_{k' | k_c} {\rm d}k'  \!\!&=\!\!  \int_{0}^{k}\!\! \left[  \mathcal{L}^{<}_{k' | k_c}  \!+\!  \mathcal{D}^{<}_{k' | k_c}   \right] {\rm d}k'  \!+\! \Pi^{<}_{k|k_c}\! + \! \Pi^{SGS}_{k|k_c}\;,
\label{eq:FE_integrated_LES}
\end{align}
where the resolved-scales-only flux and the sub-grid scale flux, respectively, are defined as, 
\begin{align}
&\Pi^{<}_{k|k_c} = \int_{0}^{k}\mathcal{T}^{<}_{k' | k_c}{\rm d}k' \;,   \\
&\Pi^{SGS}_{k|k_c} = \int_{0}^{k}\mathcal{T}^{SGS}_{k' | k_c}{\rm d}k'\;,
\end{align}
and provide the free energy transfer rate from all scales larger than $k$ to all scales smaller than $k$. At the LES cutoff $k_c$, due to the conservation of nonlinear interactions ($\mathcal{T}_{k|p,q}+\mathcal{T}_{p|q,k}+\mathcal{T}_{q|k,p}=0$), the large scales flux goes to zero ($\Pi^{<}_{k_c|k_c}=0$), as it involves all possible interactions between resolved-scales-only modes. As the total free energy flux consists in the sum of the two fluxes ($\Pi_{k}=\Pi^{<}_{k|k_c}+\Pi^{SGS}_{k|k_c}$), for $k\ge k_c$, it reduces to the SGS contribution. Since the SGS flux at the scale $k_c$ represents the energy that needs to be removed globally by the LES model, it is also known as the total sub-grid dissipation.

For GyroLES to work, the correct amount of free energy has to be dissipated. This property can be checked by  matching the sub-grid flux to the scale integrated dissipation of the model,
\begin{align}
\Pi^{SGS}_{k|k_c}  \approx    \int_{0}^{k} \mathcal{M}_{k'|k_c}[a] \,  {\rm d}k'\;,
\label{flux_mod}
\end{align}
where $\mathcal{M}_{k|k_c}[a]$ represents the free energy contribution of the model ${M}_{k|k_c}[a]$. This  can be satisfied through the free parameters of the model, i.e., finding the right parameter values that satisfy the previous relation. This is indeed a tendency that is recovered implicitly through the dynamic procedure in Eq.~\eqref{eq:c_x_y}.

\begin{figure}[!h]
\begin{center}
\includegraphics[width= 0.9\columnwidth]{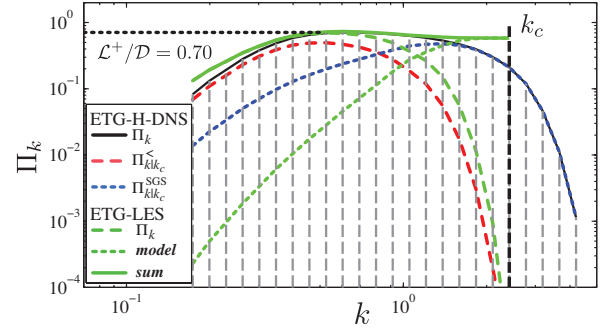}
\end{center}
\caption{Free energy flux and components for selected ETG runs. The saturation value for the flux is given as $\mathcal L^+/\mathcal D$, were $\mathcal L^+$ represents the total source of all linear terms, and $k_c$ is the LES cutoff.}
\label{fig:flux_ETG}
\end{figure}

Looking at the  ETG dataset from the perspective of the nonlinear  fluxes in Fig.~\ref{fig:flux_ETG}, we observe that the ETG-LES flux ($\Pi_{k}$, dashed-green line) seems to  match   well the largest scale $k_c$ filtered  DNS flux ($\Pi^{<}_{k|k_c}$, dashed-red line). Since both the filter DNS and the LES runs do not have any information above $k_c$, the flux goes to zero on this surface. This is also the point where the DNS flux  ($\Pi_{k}$, solid-black line)  and the SGS flux ($\Pi^{SGS}_{k|k_c}$, dotted-blue line)  have the same value, although, the SGS flux tends to make the dominant  contributions long before that point.

It is interesting to consider the scale integrated contribution of the LES model (dotted-green line). Slowly increasing in amplitude from the large scales, it saturates at the level of the maximal value of the DNS flux, without decreasing in value. This is due to the fact that the model amplitude is always positive. Looking together at the LES flux and model contributions we obtain an effective flux (solid-green line)  quickly reaches and remains at  the DNS saturation value, $\mathcal L^+/\mathcal D = 0.70$, were $\mathcal L^+$ represents the total source of all linear terms.  
This is not that surprising as the implicitly assumption of an infinite inertial range is incorporated into the model. This assumption is the reason behind the spectral slope quality of the LES runs compared to the DNS ones. 

%%%%%%%%%%
%%%%%%%%%%%%%%%%%%%%%%%%%%%%%%%%%%%%%%%%%%%%%%%%%%%%%%%%%%%%
%%%%%%%%%%%%%%%%%%%%%%%%%%%%%%%%%%%%%%%%%%%%%%%%%%%%%%%%%%%%
%
\subsubsection{Sub-grid scale locality}

It is important to remark that in the definition of the sub-grid flux, the transfer of energy from scales below $k_c$ to scales above $k_c$ does not tell us if the contribution to the flux arises primarily from scales close in value to $k_c$ or from scales with much smaller wavenumber. It also does not tell us, independently from where the energy comes from, towards which scales is the energy primarily distributed. This is very important for an application of  GyroLES models. 
Indeed, GyroLES models rely on the locality  of interactions assumption between resolved and sub-grid scales. In order to further investigate that assumptions, we will consider the classical locality    ultraviolet (UV) and infrared (IR)  functions, introduced by Kraichnan~\cite{kraichnan59} and recently applied to gyrokinetics~\cite{teaca13}, for the  SGS flux,
\begin{align}
\Pi^{SGS | UV}_{k_p | k_c} = &\int_{0}^{k_c} \!\!\mbox{d}k  \bigg{[}{\iint_{k_p}^{\infty}\!\!\! \mathcal T_{k|p,q}\, \mbox{d}p\,  \mbox{d}q}\ +\nonumber \\
2&{\int_{0}^{k_p}\!\!\!\! \int_{k_p}^{\infty} \mathcal T_{k|p,q}\, \mbox{d}p\,  \mbox{d}q} \bigg{]}    \quad {\rm with} \quad  k_p \ge k_c\,, 
\end{align}
\begin{align}
\Pi^{SGS | IR}_{k_p | k_c}  = &\int_{k_c}^{\infty} \!\! \mbox{d}k \bigg{[}{\iint_{0}^{k_p} \!\!\! \mathcal T_{k|p,q}\, \mbox{d}p\,  \mbox{d}q}\ +\nonumber \\
2&{\int_{0}^{k_p}\!\!\!\! \int_{k_p}^{\infty} \!\!\!  \mathcal T_{k|p,q}\, \mbox{d}p\,  \mbox{d}q}\bigg{]}    \quad {\rm with}  \quad k_p \le  k_c\,.
\label{eq:IR_UV}
\end{align}
It is important to differentiate between the locality of the energy cascade,  one structure giving energy to a similar size structure (as discussed in the shell-to-shell transfer section),  from the locality of interactions captured by the locality functions, where the mediator of the energetic interaction is also considered.  

The meaning of the locality functions is the following. The UV locality function represents the energy flux across $k_c$ caused by nonlinear interactions that involve at least one scale above $k_p$ (with $k_c \le k_p$).  In this case, a significant contribution would mean that the energy flux depends on the smallest scales and therefore on the type of collisions. However, due  to the resolutions employed and the LES cutoffs considered here, the UV locality information for the SGS flux is hard to  determined numerically. By definition, we are interested in seeing where the energy is being transferred across $k_c$, requiring the contributions to decrease fast as to ensure a high level of separation between the resolved and unresolved scales. As the range of scales past $k_c$ is limited and the amplitude of fluctuations are strongly damped, the UV locality information tends to be highly local.  On the other hand,  the IR locality information of the SGS flux is much more interesting to us, as it indicates the dependance of the unresolved scales to the information contained in the larger, resolved scales. 
It represents the energy flux across $k_c$ due to nonlinear interaction that involve at least one scale below $k_p$ (with $k_p \le  k_c$). Therefore,  if  there is a significant contribution, it implies that there is strong interaction with the largest scales and thus, a dependence of the type of instability that drives the system. This would imply that good GyroLES models should depend on the type of instability, and therefore, their universality could be questioned.

\begin{figure}[!htb]
\begin{center}
\includegraphics[width= 1.2\columnwidth]{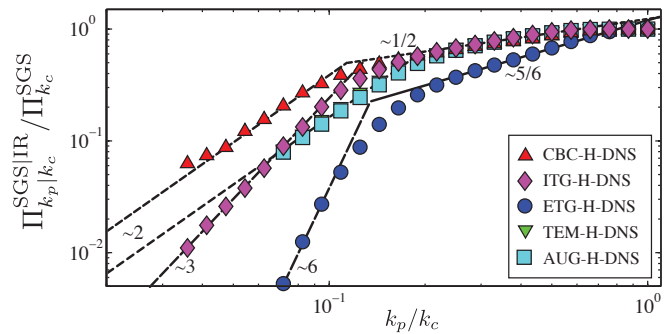}
\end{center}
\caption{IR locality functions for the SGS flux contribution. The value of $k_c$ is taken to be half the largest $k$ available for the largest DNS cases for each case. While $k_p$ wavenumbers differ from case to case, they are taken as a geometric progression with a common ratio of $2^{1/5}$.}
\label{fig_IRlocality}
\end{figure}

In Fig.~\ref{fig_IRlocality}, we plot the SGS IR locality functions normalized to the value of the flux through $k_c$. For $k_p=k_c$, as the locality functions recover the value of the flux, we obtain a unity value for this ratio. Increasing the separation between $k_p\le k_c$ and $k_c$ removes interactions from bringing contributions to the flux and as such, the ratio plotted decreases in value. The rate of this decrease gives us the assessment of the SGS flux locality. We also plot a series of slopes and their values. Except for the $(k_p/k_c)^{5/6}$ exponent value, which has a theoretical interpretation~\cite{teaca13} and is considered here as a reference, all other slopes are based on numerical observations and are given simply as a way to help us understand the results. For all runs, the fist two points smaller that $k_p/k_c=1$ have a slope close to one, as the last few physical scales tend not to be fully represented. This is just a negligible artifact, arising from the small value ($2^{1/5}$) of the common ratio of the wavenumber geometric progression, coupled with the discretization of the wavenumber space. The $2^{1/5}$ selection is taken to emphasize any slopes that might arise for the locality functions.   

Except for the ETG case, which seems to recover a $5/6$ scaling~\cite{teaca13_2}, 
all other runs have a stronger nonlocal behavior, reaching a $1/2$ slope. While this increased nonlocal tendency might be a factor to be considered for the LES modelling of the sub-grid terms (it can affect the ratio of the test filter wavenumber $k_t$ compared to the cutoff $k_c$ in the dynamical procedure), it is by no means something to  worry about. In fact, as the probe wavenumber $k_p$ starts to enter the large scale range, the locality slopes accentuate drastically (the shallowest being $\sim 2$, much more than the $5/6$ scaling~\cite{teaca13_2}). By the very nature of the driving instabilities, the locality of interaction tends to increase in locality. This might be a result of entering the driving range, a range dominated by the damped eigenmodes dynamics~\cite{hatch11}, were the cascade itself tends to be week, well below its nonlinear saturation value. Regardless of the cause, this accentuation of locality at low $k$ helps mitigate any instability dependent physics and, even if this slope is expected to be instability dependent, the high values of the exponents ensures an effective universality of the SGS modelling and validates the GyroLES approach.  

%%%%%%%%%%
%%%%%%%%%%

%%%%%%%%%%%%%%%%%%%%%%%%%%%%%%%%%%%%%%%%%%%%%%%%%%%%%%%%%%%%
%%%%%%%%%%%%%%%%%%%%%%%%%%%%%%%%%%%%%%%%%%%%%%%%%%%%%%%%%%%%
%  Conclusions
%%%%%%%%%%%%%%%%%%%%%%%%%%%%%%%%%%%%%%%%%%%%%%%%%%%%%%%%%%%%
\section{Conclusions} \label{sec:conclusion}

Via the application of the LES method to gyrokinetic turbulence driven by different kinds of microinstabilities (ITG, ETG, and TEM), two general improvements are obtained. First, the computational cost of the simulations can be considerably reduced, and second, the physical elimination of the free energy  accumulation at small perpendicular scales helps to extract the correct power law exponents. These two effects are of help in pursuing direct comparisons between numerical simulations and experiments.
 
From the study of the free energy and density fluctuation spectra, it is found that the LES method provides systematically better indication of the existence of power laws than DNS simulations used. For some cases, in order to acquire a similar accuracy as for the LES runs, DNS simulations using at least double the resolution in each of the perpendicular directions have to be performed.

The reasons for the successful implementation of the sub-grid model in the gyrokinetic LES simulations
can be briefly summarized as follows. The local character of the free energy transfer and interactions allows for the removal of small-scale interactions without affecting the overall behavior
at large scales. By modeling this effect correctly, as is done with the LES method in conjunction with the dynamic sub-grid procedure, one is able to have much better results than without it. This suggests that the LES method described in the present work is very helpful while computationally cheap, and should probably become a standard for a wide range of applications.

\section*{Acknowledgements}
The authors would like to thank V.~Bratanov, S.~S.~Cerri, G.~D.~Conway, and U.~Stroth for fruitful discussions. 
We gratefully acknowledge that the results in this paper have been achieved with the assistance of high performance computing of the HELIOS system  hosted at the International Fusion Energy Research Centre (IFERC) in Japan.  We thank the Wolfgang Pauli Institute in Vienna and the EURATOM-CIEMAT Association in Madrid for hosting international working group meetings on gyrokinetics that fostered our collaborations.
The research leading to these results received  funding from the European Research Council under the European Unions Sevenths Framework Programme (FP7/2007-2013) / ERC Grant Agreement No.~277870 and from the Princeton Plasma Physics Laboratory from the U.S.~Department of Energy under DOE Contract No.~DE-AC02-09CH11466.

%%%%%%%%%%%%%%%%%%%%%%%%%%%%%%%%%%%%%%%%%%%%%%%%%
%%%%%%%%%%%%%%%%%%%%%%%%%%%%%%%%%%%%%%%%%%%%%%%%%

%%%%%%%%%%%%%%%%%%%%%%%%%%%%%%%%%%%%%%%%%%%%%%%%%%%%%%%%%%%%
%%%%%%%%%%%%%%%%%%%%%%%%%%%%%%%%%%%%%%%%%%%%%%%%%%%%%%%%%%%%
\end{document}